\documentclass[11pt,a4paper]{article}
\usepackage{amsmath,amssymb}
\usepackage{latexsym}
\usepackage{graphicx}

\newcommand{\G}{{\mit\Gamma}}

\newcommand{\D}{\Delta}
\renewcommand{\d}{\delta}
\newcommand{\e}{\epsilon}

\renewcommand{\L}{\Lambda}
\renewcommand{\l}{\lambda}
\newcommand{\m}{\mu}
\newcommand{\n}{\nu}

\newcommand{\p}{\pi}
\renewcommand{\r}{\rho}
\renewcommand{\S}{\Sigma}
\newcommand{\s}{\sigma}

\newcommand{\f}{\phi}

\renewcommand{\o}{\omega}
\renewcommand{\O}{\Omega}

\newcommand{\be}{\begin{equation}}
\newcommand{\ee}{\end{equation}}
\newcommand{\bea}{\begin{eqnarray}}
\newcommand{\eea}{\end{eqnarray}}

\begin{document}


\begin{center}
\bf{\large  Cosmological Constant in a  Quantum Gravity Theory for a Piecewise-linear Spacetime}
\end{center}

\bigskip
\begin{center}
Aleksandar MIKOVI\'C$\,^{1,2}$ and Marko VOJINOVI\'C$\,^{3}$\\

\bigskip
$\,^{1}${\it Departamento de Matem\'atica  \\
Universidade Lus\'ofona de Humanidades e Tecnologias\\
Av. do Campo Grande, 376, 1749-024 Lisboa, Portugal}\\

\bigskip
$\,^{2}${\it Grupo de F\'isica Matem\'atica, \\
Faculdade de Ci\^ encias da Universidade de Lisboa,\\
Campo Grande, Edif\'icio C6, 1749-016  Lisboa, Portugal}\\

\bigskip
$\,^{3}${\it Institute of Physics, University of Belgrade, \\
Pregrevica 118, 11080 Belgrade, Serbia}\\

\end{center}

\centerline{E-mail: amikovic@ulusofona.pt, vmarko@ipb.ac.rs}

\bigskip
\bigskip

\begin{abstract}
\noindent\small{We study the quantum contributions to the classical cosmological constant in a quantum gravity theory for GR with matter on a piecewise linear spacetime corresponding to a triangulation of a smooth manifold.  We use the effective action approach and a special path-integral (PI) measure which depends on a free parameter, while matter is modeled by a massive self-interacting scalar field. The effective cosmological constant (CC) is given as a sum of 3 terms: the classical CC, the quantum gravity CC and the matter CC. We show that the free parameters of the theory can be chosen such that the classical CC cancels the matter CC so that the effective CC is given by the QG CC. Since the value of the quantum gravity CC is determined by the PI measure only, the PI measure parameter can be chosen such that the effective CC gives the observed value. This is equivalent to the statement that the experimentally observed CC value belongs to the spectrum of the CC operator in this QG theory.}
\end{abstract}

\bigskip
\bigskip
\noindent{\bf{1. Introduction}}

\bigskip
\noindent The cosmological constant problem, for a review see \cite{cc}, is the problem of explaining the presently observed value of the cosmological constant (CC) within a quantum theory of matter and gravitation. In any quantum gravity (QG) theory there should be a natural length scale, which is the Planck length $l_P \approx 10^{-35}\, m$. Consequently, there should be a quantum correction to the classical value of CC of order $l_P^{-2}$. However, this correction is $10^{122}$ times larger from the observed value \cite{cco}, and the problem is to explain this huge discrepancy. 

According to Polchinski, the CC problem in a QG theory has two parts \cite{cc}. The first part is to demonstrate that the observed CC value is in the CC spectrum. The second part is to explain why that particular value is selected. Hence the first step is to find out what is the CC spectrum for a given QG theory. In string theory the CC spectrum is discrete \cite{bp} and it includes positive values \cite{kklt}. However, the difficulty is to show rigorously that the observed CC value belongs to the spectrum, although it is plausible that the CC spectrum may be sufficiently dense around zero  \cite{cc,bp}. The second part of the problem is answered by using the landscape idea combined with the antrophic argument. In other QG theories, like loop quantum gravity \cite{lqg}, spin foams \cite{sfm} and casual dynamical triangulations \cite{cdt} it is still not known what is the spectrum of the cosmological constant.

Recently, a generalization of spin-foam (SF) models of QG was proposed, under the name of spin-cube (SC) models \cite{mv2p,scube}. The SC models were proposed in order to solve the two key problems of SF models: obtaining the correct classical limit and enabling the coupling of fermionic matter. This is achieved by introducing the edge lengths for a given triangulation of the spacetime as independent variables and a constraint which relates the spins for the triangles with the corresponding triangle areas. A class of spin-cube models was constructed such that it is equivalent to a Regge gravity state-sum model, which has general relativity (GR) as its classical limit, see \cite{scube}. 

A systematic study of the semiclassical approximation for the Regge state-sum models was started in \cite{amr} by using the effective action approach. A fundamental new assumption was made in \cite{amr}, and that is the hypothesis that the spacetime is described by a piecewise linear (PL) manifold corresponding to a smooth manifold triangulation. Note that in the standard quantum Regge calculus approach \cite{rc}, the spacetime triangulation is an auxilliary structure which serves to regularize the path integral, so that the spacetime is still a smooth manifold and the smooth limit has to be defined for the observables. On the other hand, in the PL manifold approach, the triangulation is a fundamental short-distance structure of the spacetime, while the smooth manifold is a long-distance approximation. This is analogous to the situation in fluid dynamics where Navier-Stokes equations are an excellent approximation for the motion of a fluid at scales much larger than the size of a fluid molecule. The main advantage of the PL spacetime approach is that the metric is described by finitely many degrees of freedom (DOF), so that the path integral (PI) becomes a finite-dimensional integral. Furthermore, when the edge lengths are much larger than $l_P$ and the triangulation has a large number of 4-simplexes, then the effective action can be approximated by the effective action for a GR quantum field theory with a cutoff given by the smallest edge length in the triangulation. 

It was also shown in \cite{amr} that by an appropriate choice of the simplex weights, or equivalently, by an appropriate choice of the path-integral measure, one can obtain a naturally small CC, of the same order of magnitude as the observed value. However, the calculation in \cite{amr} did not take into account the contributions from the matter sector, and as it is well known, the perturbative matter contributions to CC are huge compared to the observed value, see \cite{cc}. In a letter \cite{ccl} we have proposed a mechanism how to preserve the small CC value from \cite{amr} when matter is included. 

In this paper we provide the details of the calculations which were used to propose the mechanism for obtaining a small CC in \cite{ccl}. We also show that this mechanism is simply the statement that the experimentally observed CC value belongs to the spectrum of the CC operator in PLQG. The proposed mechanism for a small CC consists of choosing the two free parameters in the effective CC such that the matter contribution to CC cancels the bare value of CC so that the observable CC becomes equal to the QG contribution to CC. This procedure is equivalent to the statement that the experimentally observed CC value can be fitted in the spectrum of the CC operator in PLQG by fixing the two free parameters. Hence the first part of the CC problem is solved in PLQG. The second part of the CC problem, i.e. why is that particular value choosen, will not be addressed in this paper, because it is a more difficult problem and beyond the scope of this paper.

First we study the case of pure GR with a non-zero CC term and then we study the case with matter. We will show that the effective CC is a sum of 3 terms: the classical CC, the quantum gravity CC and the matter CC. Since the observations can only measure the sum of these 3 terms, we will show that it is possible to choose the classical CC to be equal to the negative value of the matter CC. Hence the effective CC will be given by the quantum gravity CC, which is determined by the PI measure. Since the PI measure depends on a free parameter, this parameter can be chosen such that the effective CC gives the observed value. 

It should be stressed that our approach to the problem of CC differs from the traditional QFT approach in one important aspect. Namely, in QFT the amount of matter quantum contributions to CC depends on the renormalization scale, and in order to guarantee the smallness of the observed CC one needs a mechanism to cancel this matter contribution for every choice of the renormalization scale (this requirement is usually referred to as naturalness). Our approach, however, contains a finite number of fundamental degrees of freedom describing gravity and matter. As a consequence, the matter contribution to CC is finite. In the QFT approach, which assumes a smooth spacetime, the matter contribution would be infinitely large, since the sum over all loop diagrams does not converge due to infinitely many degrees of freedom. Therefore, it is not necessary to require naturalness in our approach. 

In section 2 we study the effective action for the Regge state-sum model without matter and with a non-zero classical CC term in the semi-classical approximation. We will show that the effective CC is given by the first-order quantum correction because at the higher orders of perturbation theory there are no CC terms. We also derive the bounds for the validity of the semi-classical approximation, and this gives a restriction on the parameter of the PI measure. In section 3 we couple matter, and calculate the matter contribution to the effective CC by using the one-loop QFT approximation.  The one-loop matter CC depends on a cut-off scale, and this cut-off dependence can be removed by choosing the classical CC to be equal to the negative matter CC, so that the effective CC is given by the quantum gravity CC. This cancellation is also possible at higher-loop orders, which is shown in section 4, so that the effective CC is equal to the quantum gravity CC non-perturbatively. In section 5 we present our conclusions.

\bigskip
\bigskip
\noindent{\bf{2. Effective action for gravity with a cosmological constant}}

\bigskip
\noindent We are going to study the effective action for a discrete QG theory based on the Regge discretization of GR with a CC term. Let $T(M)$ be a simplicial complex associated with a triangulation of a 4-manifold $M=\S\times [0,1]$, where $\S$ is a compact smooth 3-manifold. The case when $\S$ is non-compact can be treated similarly to the compact case if we replace $\S$ with a ball $B$ in $\S$, such that outside of $B$ the edge lengths are kept fixed. We have restricted the topology of $M$ because we will consider only the semiclassical regime of QG where the notion of a quantum corrected classical trajectory makes sense. Consequently $(\S,0)$ is the initial spatial section and $(\S,1)$ is the final spatial section.

In each 4-simplex of $T(M)$ we will have a flat Lorentz-signature metric and  let $L_\e$, $\e =1,2,...,E$, be the edge lengths of $T(M)$, where $L_\e$ satisfy the triangle inequalities\footnote{In the usual Regge calculus one considers triangulations of manifolds with Euclidean-signature metrics. We will consider the Lorentzian signature case, so that the triangle inequalities apply only to space-like triangles. Therefore we will use only the triangulations where all the triangles are spacelike so that an edge-length is a positive square root of the Lorentz-invariant square distance of the corresponding spacetime interval. Hence an edge length is invariant under the Lorentz transformations and it is different from the spatial distance of the corresponding spacetime interval.}. The path integral of this theory, also known as the state sum, is given by the following integral
\be Z = \int_{ D_{E}} \, \mu (L) \, d^E L \,\exp \left( i{S}_{Rc} (L)/l_P^2\right) \,, \label{crss}\ee
where 
$D_E$ is a subset of ${\bf R}_+^E$ where the triangle inequalities hold and
\be {S}_{Rc} = -\sum_{\D=1}^F  A_\D (L) \theta_\D (L) + \Lambda_c\, V_4 (L) \,, \ee
is the Regge action corresponding to the Einstein-Hilbert action with the CC term, see \cite{rc}. $A_\D$ is the area of a triangle $\D$, $\theta_\D$ is the deficit angle and $V_4$ is the 4-volume of $T(M)$. The Planck length $l_P$ is given by $l_P^2 = G_N \hbar$, where $G_N$ is the Newton constant. We will also introduce a classical CC length scale $L_c$ such that
\be \Lambda_c = \pm \frac{1}{2 L_c^2} \,.\ee

We will choose the PI measure $\m(L)$ as
\be \m(L) = \exp\left(-V_4 (L)/L_0^4 \right)\,,\label{ccm}\ee
where $L_0$ is a new length scale. This type of measure ensures the finiteness of $Z$ and generates a small quantum correction to the classical CC when $\Lambda_c =0$ and $L_0 \gg l_P$, see \cite{amr}. This is also the simplest local measure which allows a perturbative effective action for large $L_\e$ and which is manifestly diffeomorphism invariant in the smooth limit ($E\to\infty$), see \cite{amr}. We have to stress that we will never take the limit $E\to\infty$, since we are postulating that the spacetime triangulation is physical, so that $E$ is fixed and the spacetime is given by the piecewise linear (PL) manifold $T(M)$. We will also assume that $E$ is a large number, i.e. $E\gg 1$, so that $T(M)$ looks like the smooth manifold $M$.

The quantum effective action $\G(L)$ associated to the theory defined by the path integral (\ref{crss}) is determined by the following integro-differential equation 
\be e^{i\G (L)/l_P^2} = \int_{D_E (L)} \,  \mu (L + l)\,d^E l \, \exp \left( i S_{Rc} (L+l)/l_P^2 - i\sum_{\e=1}^E \frac{\partial\G}{\partial L_\e }\,l_\e /l_P^2 \right)\,,\label{hde}\ee
where $D_E (L)$ is a subset of ${\bf R}^E$ obtained by translating the region $D_E$ by the vector $-L$ \cite{amr}.

When $L\to (\infty)^E$, then $D_E (L) \to {\bf R}^E$, and we can assume that the perturbative solution of (\ref{hde}) will be very-well approximated by the perturbative solution of the equation
\be e^{i\G (L)/l_P^2} = \int_{{\bf R}^E} \,\,d^E l \, \exp \left( i \bar S_{Rc} (L+l)/l_P^2 - i\sum_{\e=1}^E \frac{\partial\G}{\partial L_\e }\,l_\e /l_P^2 \right)\,,\label{pe}\ee
where
\be \bar S_{Rc}(L) = S_{Rc}(L) + il_P^2 V_4 (L) /L_0^4 \,.\ee
This assumption is based on the results of \cite{amr}, where it was shown that this is true for the exponentially damped PI measures. 

The perturbative solution of (\ref{pe}) can be written as
\be \G = \bar S + l_P^2 \bar\G_1 + l_P^4 \bar\G_2 + \cdots \,,\ee
where $\bar\G_n$ will be given by the EAD constructed for the action $\bar S_{Rc}$, see \cite{amr}. Since
\be\bar\G_n = \G_{n,0} + l_P^2 \bar\G_{n,1} + l_P^4 \bar\G_{n,2} + \cdots \,,\ee
we obtain
\be \G = S_{Rc} + l_P^2 ( -i\log\m + \G_{1,0} ) + l_P^4 ( \G_{2,0} + \bar\G_{1,1} ) + l_P^6 (\G_{3,0} + \bar\G_{1,2} +\bar\G_{2,1} ) + \cdots \,.\label{pe2}\ee
Hence
\be\G_n (L) = D_n (L) + R_n (L) \,, \ee
where $D_n$ is the contribution from the n-loop EA diagrams for the action $S_{Rc}$, while
\be R_n = Res_n \,\sum_{k=1}^{n-1} \bar D_k \,, \ee
where
\be Res_n \, f(l_P^2) = \lim_{l_P^2 \to 0}\frac{f^{(n)}(l_P^2)}{n!} \,.\ee
The $\bar D_k$ terms are defined as
\be \bar D_n (L) = D_n (L,\bar L_c^2) \,,\ee
where
\be \bar L_c^2  = L_c^2 \left( 1 + il_P^2 L_c^2 /L_{0}^4 \right)^{-1} = L_c^2 \left( 1 + il_P^2 /L_{0c}^2 \right)^{-1}\,.\label{blc}\ee

In order for the measure contributions to be perturbative, we see from (\ref{blc}) that we need $l_P / L_{0c} < 1$, which is equivalent to
\be L_0 > \sqrt{l_P L_c} \,. \label{pclz}\ee
We will study the case $L_\e > L_c$, since the perturbative analysis is simpler than in the $L_\e < L_c$ case. The large-$L$ asymptotics of $\bar\G_n (L)$ functions can be determined from
\be  S_n (L) = O(L^{4-n}) /  L_c^2  \,,\ee
and the formula for the EA diagrams, see (\ref{eadf}). Consequently, for $n>1$
\be D_{n} (L) = O\left(\left(L_c^2 / L^4 \right)^{n-1}\right) \,,\label{dna}\ee
where the $O$ notation is defined as
\be f(L) = O(L^a) \Leftrightarrow f(\lambda L ) \approx \lambda^a g(L) \ee
when $\lambda\to\infty$. Since
\be \bar\G_n (L)  = D_n (L,\bar L_c^2) \,,\ee
we obtain
\be \bar\G_{n} (L) = O\left(\left(\bar L_c^2 / L^4 \right)^{n-1}\right) \,.\label{bgna}\ee

The asymptotics (\ref{dna}) can be derived by considering the one-dimensional ($E=1$) toy model
\be S_{Rc} = \left(L^2 + \frac{L^4}{L_c^2}\right)\theta (L) \,, \ee
where $\theta (L)$ is a homogeneous $C^\infty$ function of degree zero. Consequently
\be D_n (L) = \sum_{l\in{\bf N}} c_{nl}\,(G(L))^{k_l}\, S_{n_1}(L) \cdots S_{n_l}(L)\,, \label{eadf}\ee
where $G = 1/S_{Rc}''$, $S_n = S_{Rc}^{(n)}/n!$, $k_l$ is the number of edges of an $n$-loop EA graph with $l$ vertices and $c_{nl}$ are numerical factors.

The asymptotics (\ref{dna}) implies that there are no $O(L^4)$ terms in $D_n(L)$, and hence $D_n (L)$ cannot contribute to the effective CC. This also happens for the $R_n$ terms, which can be seen from the toy model, where
\be \bar S''_{Rc} = \theta_1 (L) [ 1 + (L^2 /\bar L_c^2 ) \,\theta_2 (L)] \,,\ee
and $\theta_k$ are homogeneous functions of degree zero. Consequently
\be \log \bar S''_{Rc} = \log (L^2 /\bar L_c^2) + \log \theta_1 (L) + \log \left[ 1 + O(\bar L_c^2 /L^2)\right]  \,, \ee
while from (\ref{bgna}) it follows that
\be R_n (L) = O((L_{0c}^2 )^{-n+1}) + O(L^{-2} (L_{0c}^2)^{-n+2}) + O(L^{-4} (L_{0c}^2)^{-n+3}) + \cdots \,.\ee
We then obtain
\be \G_1 = O(L^4/L_0^4) + \log O(L^2/L_c^2) + \log\theta_1 (L) + O(L_c^2 /L^2) \,,\ee
and
\be \G_n = D_n + R_n = O((L_c^2 /L^4)^{n-1}) + L_{0c}^{2-2n} \, O(L_c^2 /L^2) = L_{0c}^{2-2n} \, O(L_c^2 /L^2)\,.\ee
Note that we have discarded the constant pieces in $\G_n (L)$.

Hence there are no $O(L^4)$ terms in $\G_n$ for $n > 1$ and therefore the effective cosmological constant will be determined by the $\log\m$ term, so that
\be \Lambda_g =\Lambda_c + \Lambda_\m = \pm \frac{1}{2 L_c^2}  \pm \frac{l_P^2}{2 L_0^4} \,. \label{qgcc}\ee
The formula (\ref{qgcc}) follows from the physical effective action, which is defined as
\be S_{eff} = (Re\,\G \pm Im\,\G)/G_N \,. \label{phea}\ee
We have used in (\ref{phea}) the QG Wick rotation 
\be \G \to Re\,\G \pm Im\,\G \,, \label{wr}\ee
in order to make the effective action a real function, since the solutions of the EA equation are complex\footnote{In QFT, the Wick rotation $t \to it$, where $t$ is the time coordinate in a flat spacetime, transforms the EA equation into a real integro-differential equation and the Minkowski metric in $\G$ becomes a Euclidean metric. Consequently the solutions $\G_{it}$ of the Wick-rotated EA equation are real, so that when one substitutes the Euclidean metric in $\G_{it}$ with a Minkowski metric, one obtains a real $\G$. This is equivalent to performing the transformation (\ref{wr}). In the QG case there is no analogue of the coordinate $t$, and the transformation (\ref{wr}) is a coordinate-free analogue of the Wick rotation.}, see \cite{mvea,scube}. The sign ambiguity in (\ref{phea}) will be fixed by requiring that $\Lambda_\m$ is positive, see the next section.

Note that the condition (\ref{pclz}) and $L_\e > L_c$ ensure that the effective action is semiclassical, which implies that the quantum corrections to the classical action will be small for 
\be L_0 \gg \sqrt{l_P L_c}\,. \label{scc}\ee 
In this case
\be  |S_{Rc}|/l_P^2 \gg |\G_1| = |\log\m - \frac{1}{2} Tr\,\log S''_{Rc}| \,,\label{sc}\ee
and
\be |\G_n | \gg l_P^2 | \G_{n+1}| \,,\label{scn}\ee
for all $n$.

Also note that the effective action will remain semiclassical if $L_c$ is large and $ L_\e < L_c$, but in this case we need $L_\e \gg l_P$ in addition to the condition (\ref{pclz}). This can be seen from the asymptotics of $\bar\G_n (L)$ terms when $L_\e < L_c$, since
\be \log\bar S''(L) = \log\theta_1 (L) + \log\left[1 + O(L^2 /\bar L_c^2)\right]\label{asa}\ee
and
\be \bar\G_{n+1}(L) = O(1/L^{2n})\left[1 + O(L^2 /\bar L_c^2)\right]\,.\label{asb}\ee

The asymptotics (\ref{asa}) and (\ref{asb}) imply that we may have terms of $O(l_P^{2n})$ for any $n$ contributing to $\L_g$. However, since $\L_g$ is a constant, i.e. it is independent of $L$, and given that we showed for $L_\e > L_c$ that there are no such terms in $\L_g$, see equation (\ref{qgcc}), then this implies that in the case $L_\e < L_c$ the sum of $O(l_P^{2n})$ terms must be zero. Hence one obtains the same formula for $\L_g$ as (\ref{qgcc}).

We will consider an edge length $L_\e$ to be large if $L_\e \gg l_P$, so that a triangulation will have large edge lengths if
\be L_\e \ge L_K \gg l_P \,,\label{qftk}\ee
where $L_K$ is the minimal edge length. The length $L_K$ will serve as a QFT cutoff in the smooth-manifold approximation of the effective action.

\bigskip
\bigskip
\noindent{\bf{3. Effective action for gravity with a scalar field}}

\bigskip
\noindent In order to see what is the effect of matter on the value of CC we will consider a scalar field $\f$ on a 4-manifold $M$ with a metric $g$ such that the scalar-field action is given by
\be S_s (g,\f) = \frac{1}{2}\int_M d^4 x \sqrt{|g|}\left[g^{\m\n}\,\partial_\m \f \,\partial_\n \f -  U(\f) \right]\,, \label{sca}\ee
where $U(\f)$ is a polynomial of the degree greater or equal than 2.

When the metric $g$ is non-dynamical, the EOM of (\ref{sca}) are invariant under the constant shifts of the potential $U$. However, we know that the metric is dynamical, so that the constant shifts in $U$ will give contributions to the cosmological constant term. These classical shifts of the potential will affect the value of $\Lambda_c$, so that we will assume that $\Lambda_c \ne 0$.

On $T(M)$ the action (\ref{sca}) becomes
\be S_{Rs} = \frac{1}{2}\sum_\s V_\s (L)  \sum_{k,l} g^{kl}_\s (L)\,  \f'_k \, \f'_l - \frac{1}{2}\sum_\pi V_\pi^* (L)\, U( \f_\pi) \,,\ee
where $g^{kl}_\s$ is the inverse matrix of the metric in a 4-simplex $\s$
\be g_{kl}^{(\s)} = \frac{ L_{0k}^2 + L_{0l}^2 -L_{kl}^2}{L_{0k} \, L_{0l}} \,,\label{cm}\ee
$ \f'_k = (\f_{\pi_k} - \f_{\pi_0})/L_{0k}$ and $V^*_\pi$ is the volume of the dual cell for a vertex point $\pi$ of $T(M)$, see \cite{rc}\footnote{In \cite{rc} the cell metric is given by the numerator of (\ref{cm}). The denominator in (\ref{cm}) appears after performing a coordinate transformation in $\s$, and we did this in order to obtain a dimensionless expression for the cell metric.}.

The quantum corrections due to gravity and matter fluctuations can be described by the effective action based on the classical action
\be S(L,\f) = \frac{1}{G_N}S_{Rc}(L) + S_{Rs} (L,\f) \,.\label{cla}\ee
Since
\be S(L,\f)/\hbar = S_{Rc}(L)/l_P^2 + G_N S_{Rs} (L,\f)/l_P^2 = S_{Rm}(L,\f)/l_P^2 \ee
the EA equation becomes 
\bea e^{i\G (L,\f)/l_P^2} &=& \int_{D_E (L)} d^E l \,\int_{{\bf R}^V}\prod_\pi d\chi_\pi \exp \Big{[} i \bar S_{Rm} (L+l, \f + \chi)/l_P^2 \cr
&-&i\sum_\e \frac{\partial\G}{\partial L_\e }\,l_\e /l_P^2 -i\sum_\p \frac{\partial\G}{\partial \f_\p }\,\chi_\p /l_P^2 \Big{]}\,,\label{mpe}\eea
where $\bar S_{Rm} = \bar S_{Rc} + G_N S_{Rs} (L,\f)$.

Since we are using an exponentially damped PI measure for the $L$ variables, we can use the approximation $D_E(L) \approx {\bf R}^E$ when $L_\e \to \infty$, see \cite{amr}. We can then solve (\ref{mpe}) perturbatively in $l_P^2$ by using the EA diagrams for the action $\bar S_{Rm}$.

It is convenient to introduce a dimensionless field $ \sqrt{G_N}\,\f$, so that
$\sqrt{G_N}\,\f \to \f$ and $S_{Rm} = S_{Rc} + S_{Rs}$. The perturbative solution will be given by
\be \G(L,\f) = S_{Rm} (L,\f) + l_P^2 \G_{1}(L,\f) + l_P^4 \G_{2}(L,\f) + \cdots\,, \label{pme}\ee
where $\G_{n}$ are given by the EA diagrams corrected by the measure contributions, see section 2.
It is not difficult to see that
\be \G (L,\f) = \G_{g}(L) + \G_m (L,\f) \,, \label{lfd}\ee
and that for constant $\f$ configurations
\be \G_m (L,\f) = V_4 (L)\, U_{eff}(\f) \,.\label{cf}\ee

We expect that the expansion (\ref{pme}) will be semiclassical for $L\gg l_P$ and $\f \ll 1$. This can be verified by studying the one-dimensional ($E=1$) toy model for the potential
\be U(\f) = \frac{\o^2}{2} \f^2 + \frac{\lambda}{4!}\f^4 \,,\label{sp}\ee
where $\hbar\o = m$ is the matter field mass and $\lambda$ is the matter self-interaction coupling constant.
The toy-model classical action can be taken to be
\be S_{Rm} (L,\f) = \left(L^2 + \frac{L^4}{L_c^2}\right)\theta (L) + L^2 \left[\f^2 + \frac{L^2}{L_m^2 } (\f^2 + a\f^4)\right]\theta(L) \,, \ee
where $L_m = 1 /\o$, $\lambda/4! = a/L_m^2$ and the PI measure $\m = \exp(-L^4 /L_0^4)$.

The first-order quantum correction to the classical action (\ref{cla}) is determined by
\be \G_{1} =  i\frac{V_4}{L_0^4} + \frac{i}{2} Tr \log \begin{pmatrix} 
 S_{LL} &  S_{L\f} \\ 
 S_{L\f} &  S_{\f\f}
\end{pmatrix} 
\,,\ee
where $S_{xy}$ are the submatrices of the Hessian matrix for $S_{Rm}$. Since 
\be S_{LL} = O(L^2)\,,\quad S_{L\f} = O(L^3)O(\f)\,,\quad S_{\f\f} = O(L^4)[1 +  O(\f^2)] \,,\ee
for $L$ large, then
\be \G_1 = i\frac{V_4(L)}{L_0^4} + \frac{i}{2} Tr \log S_{LL} + \frac{i}{2} Tr \log S_{\f\f} + O(\f^2) \,.\label{olm}\ee

The first term in (\ref{olm}) is the QG correction to the classical CC, while the matter sector will give a quantum correction to CC from the third term. This can be seen by considering the smooth manifold approximation, i.e. when $E \gg 1$. In this case the third term in (\ref{olm}) can be calculated by using the continuum approximation
\be S_{Rs}(L,\f) \approx S_s (g,\f) \,,\ee
and the corresponding QFT in curved spacetime.  

Let us consider an edge-length configuration which satisfies (\ref{qftk}). 
The condition (\ref{qftk}) ensures that the QG corrections are small and if $L_K \ll L_m$, we can calculate $Tr\,\log S_{\f\f}$ by using the Feynman diagrams for $S_s$ with the UV momentum cutoff $\hbar /L_K = \hbar K$. Consequently
the corresponding CC contribution will be given by the flat space vacuum energy density, since
\be   Tr \log S_{\f\f}\big{|}_{\f=0} \approx V_M \int_0^K k^3 \,dk \,\log (k^2 + \o^2 ) + \O_m(R,K) \equiv \d\G_1 (L) \,,\ee
and
\bea \O_m (R,K) &=& a_1 K^2 \int_M d^4 x \sqrt{|g|}\,R \cr
&+&  \log(K/\o)\, \int_M d^4 x \sqrt{|g|}\left[ a_2 R^2 + a_3 R^{\m\n}R_{\m\n} +a_4 R^{\m\n\r\s} R_{\m\n\r\s}+ a_5\nabla^2 R \right]\cr &+& O\left(L_K^2 /L^{2}\right) \,,\eea
where $a_k$ are constants, see \cite{bd}. Therefore the only $O(L^4)$ term in $\d\G_1$ is
\be c_1 V_M \, K^4 \log\left(K/\o \right)  = c_1  \, \frac{V_M}{L_{K}^4} \log(L_m /L_{K}) \,,\ee
where $c_1$ is a numerical constant.

The physical effective action is given by the formula (\ref{phea}), so that the one-loop CC is given by
\be \Lambda_1 = \pm \frac{1}{2L_c^{2}} + \Lambda_\m +  c_1 \, \frac{l_P^2}{2 L_{K}^4} \log(K/\o) \,, \ee
where $c_1$ is a numerical constant of $O(1)$. We can write this as
\be \Lambda_1 = \Lambda_{\m} + \Lambda_c + \Lambda_{m} \,,\ee
and it is not difficult to see that the higher-loop matter contributions to CC will preserve this structure, due to (\ref{lfd}) and (\ref{cf}). In the next section we will give a detailed demonstration of this. Consequently
\be \Lambda = \Lambda_\m + \Lambda_c + \Lambda_m \,,\label{ccf}\ee
where
\be \Lambda_m \approx \frac{l_P^2}{L_{K}^4} f(\bar\lambda , K^2 /\omega^2 ) \,,\label{lm}\ee
$\bar\lambda = \lambda\, l_P^2$ and $f(x,y)$ is a $C^\infty$ function, see the next section. 

We can then choose the free parameter $L_c$ such that 
\be \Lambda_c + \Lambda_m = 0 \,,\label{canc}\ee 
so that
\be \Lambda = \Lambda_\m = \frac{l_P^2}{2L_0^4} \,.\label{laf}\ee
Note that $\Lambda_\m > 0$ if we choose the $+$ sign in (\ref{phea}).

By taking $L_0 \approx 10^{-5}\,m$ we obtain the observed value of CC, which is
\be l_P^2 \Lambda_\m \approx 10^{-122}\,.\ee
Note that $L_0 \approx 10^{-5}\,m$ satisfies $L_0 \gg l_P$, which is consistent with the condition (\ref{scc}) for the validity of the semiclassical approximation. Namely, if $L_c \ge l_P$, then (\ref{scc}) implies $L_0 \gg l_P$. If $L_c < l_P$, then $L_0 \gg l_P$ is consistent with $L_0 \gg \sqrt{l_P L_c}$ since 
\be L_0 \gg l_P > \sqrt{l_P L_c}\,.\ee
This is important because the value of CC can be measured only in the semiclassical regime of a QG theory.

Note that the final expression for $\L$, eq. (\ref{laf}), is cut-off independent. This must happen because the cut off $K$ is an artefact of the QFT approximation, which also requires that the edge lengths are large. However, the exact solution of the EA equation, $\G (L)$, is defined for all allowed values of $L$ and it will not depend on $L_K$. Hence the corresponding $\L$ can only depend on the input parameters of the theory, which are $l_P$, $L_0$, $L_c$, $\l$ and $\o$. The QFT approximation together with the cut-off independence implies that the exact solution for the EA will give
\be \L = {l_P^2 \over 2 L_0^4} + \L_c + v(l_P,\l,\o) \,.\label{ecc}\ee
The free parameters $L_0$ and $\L_c$ can be determined from the the condition $\L_c +\L_m = 0$ and from the observed value of $\L$.

The formula (\ref{ecc}) allows us to choose $\L_c + \L_m$ to be a non-zero constant $C$. Then $L_0$ becomes a function of $C$, and the condition $L_0 \gg l_P$, together with the observed value of $\L$, gives a range for $C$
\be -10^{-122} < C\,l_P^2 \ll 1 -10^{-122}\,. \label{cint}\ee
Clearly, the simplest and the most natural choice is $C=0$, but any other $C$ from the interval (\ref{cint}) can be taken. The only difference is that $C\ne 0$ will give a different value for $L_0$.

Note that the formula (\ref{ecc}) also allows the choice $\L_\m + \L_m = 0$. Then one would get that $\L = \L_c$ so that $\L_c$ will be equal to the observed value of $\L$. However, the problem with this approach is that the condition
\be \frac{l_P^2}{2L_0^4} + v(l_P ,\l ,\o) = 0 \,,\ee
should determine the value of $L_0$, but we do not know the exact value of $v$. Hence we would not be able to check the consistency condition $L_0 \gg l_P$.

\bigskip
\bigskip
\noindent{\bf{4. Higher-loop matter contributions to CC}}

\bigskip
\noindent In this section we will prove the formula (\ref{lm}) for the matter contributions to CC. The matter contributions are given by the sum of $n$-loop one-particle-irreducible (1PI) QFT Feynman diagrams with no external legs and with a momentum cut-off $\hbar K$. This is because the $\f$-independent terms in the effective action are determined by the non-zero EA diagrams such that the $\f\to 0$ limit is taken in the propagator and the vertex functions. This leaves only the matter 1PI vacuum-energy diagrams.

Let $U(\f)$ be given by (\ref{sp}), then the contribution to $\Lambda_m$ of $O(\hbar^n)$ is given by the sum of $n$-loop 1PI vacuum diagrams, which we denote as $\delta_n \Lambda_m$. This sum can be represented as
$$
\d_n \L_m = \langle
\raisebox{-8pt}{\includegraphics[bb=0 0 106 23]{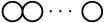}}
\rangle_n + \langle
\raisebox{-23pt}{\includegraphics[bb=0 0 88 54]{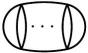}}
\rangle_n + 
$$
\be  + \langle
\raisebox{-31pt}{\includegraphics[bb=0 0 70 70]{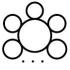}}
\rangle_n + \langle
\raisebox{-35pt}{\includegraphics[bb=0 0 79 77]{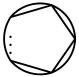}}
\rangle_n + \cdots \,,\ee
where the chain graphs appear for $n\ge 2$, watermelon graphs appear for $n\ge 3$, flower and polygon-in-a-circle graphs appear for $n\ge 4$, and so on.

We would like to determine the large-$K$ behavior of these graphs. This asymptotics is generically given by $O(K^D)$, where $D$ is the degree of the superficial divergence of the graph. However, there are exceptions, and we will show that this happens in the case of flower graphs.

The 2-loop matter contribution to CC is given by the chain graph
\be\delta_2 \Lambda_{m} = c_2 \, \lambda \, l_P^4  \left(\int_0^K \frac{k^3 dk}{k^2 + \o^2}\right)^2 \approx c_2 \,\lambda \, l_P^4 \, K^4 = c_2 \, \frac{l_P^4}{L_\lambda^{2}L_K^{4}} \,. \ee
since $K \gg \omega $. This agrees with $D=4$ for the 2-loop chain graph.

At 3 loops we have the chain graph contribution
\be\d_3^C \Lambda_{m} = c_3 \, \l^2 \, l_P^6  \left(\int_0^K \frac{k^3 dk}{k^2 + \omega^2}\right)^2 \int_0^K \frac{q^3 dq}{(q^2 + \omega^2)^2 }\approx c_3 \,\lambda^2 \, l_P^6 \, K^4 \ln (K^2 / \omega^2 ) \,. \label{tca}\ee
This graph has $D=4$ and the asymptotics (\ref{tca}) is consistent with this value of $D$.

For the 3-loop melon graph we obtain
\bea \delta_3^M \Lambda_{m} &=& m_3 \, \lambda^2 \, l_P^6  \int_0^K \frac{k^3 dk}{k^2 + \o^2} \int_0^K \frac{q^3 dq}{q^2 + \omega^2 } \int_{r \le K} \frac{d^4 \vec r}{(r^2 + \omega^2) [(\vec r-\vec k-\vec q)^2 + \omega^2]}\cr &\approx& m_3 \,\lambda^2 \, l_P^6 \, K^4 \ln (K^2 / \omega^2 ) \,, \eea
which again agrees with the corresponding $D$.

At 4 loops the flower graph appears, and it gives 
\be \delta_4^F\L_m = f_3 \, \lambda^3 \, l_P^6  \left(\int_0^K \frac{k^3 dk}{k^2 + \omega^2}\right)^3 \int_0^K \frac{q^3 dq}{(q^2 + \omega^2)^6 } \,.\ee
This integral has $D=4$, but its asymptotics is given by $D=6$. The reason is that the second integral is not asymptotic to $K^{-2}$ but it is asymptotic to a non-zero constant, so that
\be \delta_4^F\L_m \approx f_4 \, l_P^2 K^4 \,\bar\lambda^3 (K/\omega)^2 \,.\ee

An $n \ge 3$ chain graph gives
\bea \delta_{n}^C \L_{\f} &=& c_n \, \lambda^{n-1} \, l_P^{2n}  \left(\int_0^K \frac{k^3 dk}{k^2 + \omega^2}\right)^2 \left(\int_0^K \frac{k^3 dk}{(k^2 + \omega^2)^2 }\right)^{n-2} \cr
&\approx& c_n \,\lambda^{n-1} \, l_P^{2n} \, K^4 \left(\ln (K^2 / \omega^2 )\right)^{n-2} \,, \eea
while an $n\ge 4$ polygon graph gives
\bea \delta_{n}^P \Lambda_{\f} &=& p_n \, \lambda^{n-1} \, l_P^{2n}  \int_0^K \frac{k^3 dk}{k^2 + \omega^2} \int_0^K \frac{q^3 dq}{ q^2 + \omega^2 } \left(\int_{r \le K} \frac{d^4 \vec r}{(r^2 + \omega^2) [(\vec r-\vec k-\vec q)^2 + \omega^2]}\right)^{n-2}\cr &\approx& p_n \,\lambda^{n-1} \, l_P^{2n} \, K^{4} \left(\ln (K^2 / \omega)\right)^{n-2} \,. \eea
A flower graph gives for $n\ge 4$
\be \delta_{n}^F\Lambda_m \approx f_n \,l_P^2 K^4 \,\bar\lambda^{n-1} (K^2 /\omega^2)^{n-3} \,.\ee

As far as the other 1PI vacuum graphs are concerned, their $D$ is less than $4$, and consequently the main contribution for large $K$ is given by 
\bea \Lambda_m \approx l_P^2 \,K^4 {\Big[} &c_1& \ln (K^2 /\omega^2) + \sum_{n\ge 2} c_n {\bar\lambda}^{n-1} \left(\ln (K^2 /\omega^2)\right)^{n-2} \cr &+& \sum_{n\ge 4} d_n {\bar\lambda}^{n-1} \left( K^2 /\omega^2 \right)^{n-3} {\Big ]}\,, \eea
where $\bar\lambda = \lambda\, l_P^2 $ is dimensionless. Since $K \gg \omega$, we get
\be \Lambda_m \approx l_P^2 \,K^4  \sum_{n\ge 4} d_n {\bar\lambda}^{n-1} \left( K^2 /\omega^2 \right)^{n-3}  \,, \label{aplf}\ee
so that the flower graphs have a dominant contribution.

This expansion will be perturbative if 
\be \bar\lambda K^2 /\omega^2 < 1 \,.\ee
Since $\bar\lambda = 1/8$ and from
\be K \gg \omega \,,\ee
we get $K /\omega = 10^k$ where $ k \ge 2$. Hence $10^{2k-1} < 1$, which is not possible for $k \ge 2$. Therefore for a given $K$ we have to calculate $\L_m$ for a large number of loops in order to obtain an accurate value. 

Hence (\ref{aplf}) is a perturbative approximation of an exact non-perturbative value for $\Lambda_m$, valid for $L_K \gg l_P$. We can then write
\be  \Lambda_m \approx l_P^2 \,K^4 \,f(\bar\lambda , \, K^2 / \omega^2 ) \,, \ee
where $f$ is an analytic function. The exact value for $\L_m$ will be cut-off independent, so that
\be \L_m = v(\l,\o,l_P) \,, \label{ev}\ee
where the function $v$ will be determined by the exact solution of the EA equation.

\bigskip
\bigskip
\noindent{\bf{5. Conclusions}}

\bigskip 
\noindent We have shown that the CC in a discrete QG theory based on the Regge formulation of GR is given by the eq. (\ref{ecc}). The QG contributions to CC can be calculated explicitly, and they are given by a simple expression (\ref{qgcc}). The matter contributions to CC are well approximated by the EA loop diagrams for the matter QFT with a physical momentum cut-off $\hbar /L_K$, where $L_K \gg l_P$, see (\ref{aplf}).  This contribution cannot be calculated exactly, but it will have a definite value, see (\ref{ev}), since the effective action is defined non-perturbatively via the equation (\ref{mpe}). 

Due to the additive structure of the quantum contributions to CC (\ref{ccf}) and their functional dependence on the input parameters of the theory ($l_P, L_0, L_c, \o, \l$),
we can choose $\Lambda_c$ such that $\Lambda_c = -\Lambda_m$ so that $\Lambda = \Lambda_\m = l_P^2 /2L_0^4$, where $L_0$ is a free parameter entering the QG path-integral measure. Note that the choice $\L_c +\L_m =0$ ensures the cut-off independece of $\L$ in the QFT approximation, i.e. $\L$ should not depend on the minimal length $L_K$. By choosing $L_0 \approx 10^{-5}\,m$ we obtain the presently observed value of the CC. This value of $L_0$ is consistent in our approach, because it satisfies $L_0 \gg l_P$, which is the condition for the validity of the semiclassical approximation, see (\ref{scc}). 

According to the classification of \cite{cc}, our QG theory belongs to the class of adjustable-$\L$ theories. This means that $\L$ depends on free parameters, in our case $L_0$ and $L_c$, which are adjusted such that one obtains the observed value. This is a nontrivial task, because one has to find out the dependence of $\L$ on the free parameters for a given QG theory, and than to show that there is a consistent solution of the equation $\L = \L_0$. In our case the consistency condition is $L_0 \gg l_P$, which is clearly satisfied.

We have shown that the QG semiclassical approximation is valid for $L_K \gg l_P$. Note that $L_K$ is a scale where the QG corrections are still small and the usual perturbative QFT is still valid. From the LHC experiments we know that QFT is valid at the length scales of the order of $10^{-20}\,m$ so that $l_P \ll L_K < 10^{-20}\,m$.

It is important to emphasize that the choice (\ref{canc}) is not the same as the extreme fine tuning one needs to perform in order to obtain the observed value of CC when ignoring the QG effects, since one does not need to know the value of $\L_m$. The choice (\ref{canc}) can be understood as a way to enforce the cut-off independence of $\L$. From this point of view, one can also choose $\L_c + \L_m = C$, where the constant $C$ belongs to the interval (\ref{cint}). Whether $C=0$ is a special value or a matter of convenience, remains to be seen, but for the purposes of obtaining the desired value of CC, this issue is not important.

Our results demonstrate that there is a simple QG theory which satisfies the basic requirements for a QG theory: to be well-defined and to have a good semiclassical limit. This is achieved by assuming that the spacetime is described by a piecewise-linear manifold corresponding to a triangulation of a smooth manifold. Hence in this QG theory the spacetime triangulation is physical, and not an auxiliary structure whose purpose is to define the smooth manifold limit. A curved space QFT can be then understood as an approximation for this QG theory with finitely many degrees of freedom, which is analogous to fluid dynamics being an approximation for the molecular structure of a fluid.

Furthermore, the PL QG theory can reproduce the observed value of the CC in the sense that the observed value of the CC belongs to the spectrum of the CC operator. The QG corrections to the classical action can be calculated by using the usual QFT with a physical UV cutoff $K$, which corresponds to the minimal edge length $L_K$ in a triangulation with large edge-lengths $L_\e \gg l_P$, so that $L_K \gg l_P$. There is an upper bound on $L_K$, coming from the LHC experimets, and it is given by $L_K < 10^{-20}\,m$. An important assumption for the validity of the QFT approximation is that the number of the edge-lengths $E$ in a triangulation $T(M)$ is large, so that we can approximate the discrete sums over the cells in $T(M)$ with the integrals over the smooth manifold $M$.

\bigskip 
\bigskip
\noindent{\bf Acknowledgments}

\bigskip
\noindent This work has been partially supported by GFMUL and the Serbia-Portugal bilateral project ``Quantum Gravity and Quantum Integrable Models - 2015-2016'' number 451-03-01765/2014-09/24. AM was partially supported by the FCT project ``Geometry and Mathematical Physics'', EXCL/MAT-GEO/0222/2012. MV was also supported by the project ON171031 of the Ministry of Education, Science and Technological Development, Serbia.


\end{document}